\documentclass[prl,twocolumn,superscriptaddress,nobibnotes,letterpaper]{revtex4}

\usepackage{amsmath}
\usepackage{amssymb}
\usepackage{amsfonts}
\usepackage[usenames, dvipsnames]{color}
\usepackage{wrapfig}
\usepackage{eucal}
\usepackage{hhline}
\usepackage{threeparttable}
\usepackage{supertabular}
\usepackage{multirow}
\usepackage{tabularx}
\usepackage{mathtools}

\usepackage{graphicx}
\usepackage{tikz}

\begin{document}

\pagenumbering{arabic}

\title{Integrated optical force sensors using focusing photonic crystal arrays}\thanks{This work was published in Opt.\ Express \textbf{25}, 9196--9203 (2017).}

\normalfont

\author{Jingkun Guo}
\affiliation{Kavli Institute of Nanoscience, Delft University of Technology, Lorentzweg 1, 2628CJ Delft, The Netherlands}
\author{Richard A.\ Norte}
\affiliation{Kavli Institute of Nanoscience, Delft University of Technology, Lorentzweg 1, 2628CJ Delft, The Netherlands}
\author{S.\ Gr\"oblacher}
\email{s.groeblacher@tudelft.nl}
\affiliation{Kavli Institute of Nanoscience, Delft University of Technology, Lorentzweg 1, 2628CJ Delft, The Netherlands}


\begin{abstract}
Mechanical oscillators are at the heart of many sensor applications. Recently several groups have developed oscillators that are probed optically, fabricated from high-stress silicon nitride films. They exhibit outstanding force sensitivities of a few aN/Hz$^{1/2}$ and can also be made highly reflective, for efficient detection. The optical read-out usually requires complex experimental setups, including positioning stages and bulky cavities, making them impractical for real applications. In this paper we propose a novel way of building fully integrated all-optical force sensors based on low-loss silicon nitride mechanical resonators with a photonic crystal reflector. We can circumvent previous limitations in stability and complexity by simulating a suspended focusing photonic crystal, purely made of silicon nitride. Our design allows for an all integrated sensor, built out of a single block that integrates a full Fabry-P\'{e}rot cavity, without the need for assembly or alignment. The presented simulations will allow for a radical simplification of sensors based on high-Q silicon nitride membranes. Our results comprise, to the best of our knowledge, the first simulations of a focusing mirror made from a mechanically suspended flat membrane with subwavelength thickness. Cavity lengths between a few hundred $\mu$m and mm should be directly realizable.
\end{abstract}

\maketitle

\section{Introduction}

Over the past years, several designs of high-index contrast gratings have been investigated for various applications, like filters~\cite{Boutami2006,Wang2015a,Wang2015b}, microcavities~\cite{Li2011} or as planar alternatives to lenses~\cite{Feng2005,Fattal2010,Lu2010,Zhan2016}. Some of the designs feature membranes~\cite{Letartre2003,Mateus2004} and even incorporate tunability of the focal length~\cite{Kamali2016}. At the same time, simultaneously designing and fabricating high reflectivity and good mechanical quality in an integrated MEMS structure has been a long outstanding goal. First realization of such devices have been made by patterning a diffraction grating into a silicon nitride membrane~\cite{Stambaugh2015}. Recently several groups have focused on using 2D photonic crystal arrays~\cite{Crozier2006,Bui2012,Norte2016,Bernard2016,Chen2017}, achieving reflectivities beyond 99\%. Such devices are interesting for quantum optomechanics experiments, as they in principle allow for ground-state cooling from room temperature with potentially increased optomechanical coupling rates~\cite{Xuereb2012}. In order to combine the outstanding force sensitivity of these recent optomechanical designs~\cite{Norte2016} with the capability of building a fully integrated optical sensor, it is very desirable to design and fabricate optical cavities directly on a chip. This avoids complex infrastructure for alignment and read-out and is inherently stable, opening up to possibility to use these sensors for easy to use, plug-and-play applications. Here, we present a design of a focusing 2D photonic crystal array, that avoids using several layers of dielectrics but can be directly patterned into a single sheet of material, like highly-stressed silicon nitride, which is known for its exceptionally low mechanical dissipation. This potentially allows the realization of a high quality mechanical system, necessary to achieve state-of-the-art force sensitivity, with a direct optical readout through a Fabry-P\'{e}rot cavity, all integrated on a single chip.

\section{Design approach}
The general procedure to design our devices is as follows:\ we start with a two-dimensional photonic crystal (PC) slab -- a thin, periodic photonic structure. Using Bloch's theorem, the electro-magnetic (EM) waves in the slab can be decomposed into eigenmodes characterized by different bands and discrete wave vectors~\cite{Joannopoulos2008}. The profiles of the modes, and hence the optical properties of the slab, are governed by a single unit cell. By choosing a proper design of the unit cell an incoming mode can interfere destructively, resulting in high reflectance of the slab for a certain wavelength. In addition to controlling the amplitude reflectance, the wave also acquires a phase shift. Both the reflectivity and phase shift can be calculated using finite element method simulations with Floquet boundary conditions or rigorous coupled wave analysis (RCWA)~\cite{Moharam1981,Rumpf2011}, which can give modes and reflection properties at the same time.

In order to realize a focusing mirror, constructive interference of the reflected wave at the focal point is required. We can write the required profile of the phase shift on the reflector as
\begin{equation}
\phi_{\mathrm{tag}}(\vec x) =
\frac{2\pi}{\lambda} (f-\sqrt{f^2+|\vec x|^2}) + \phi_0,
\end{equation}
where $\lambda$ is the wavelength, $f$ is the focal length, and $\vec x$ is the location of a point on the reflector. $\phi_0$ is a constant phase shift that can be neglected. This gives rise to an ideal thin lens focusing~\cite{Fattal2010,Lu2010}. To calculate the focusing of the PC deterministically, a conventional PC is modified adiabatically such that it matches the target phase profile locally. A practical approach is to parameterize the geometry, and modulate the parameters of each single cell. Unavoidably, the periodicity is broken, however as is shown below, the structure can still retain high reflectivity, as has also been demonstrated in high-contrast photonic grating arrays~\cite{Fattal2010,Lu2010,Klemm2013}.

\begin{figure}[t]
	\centering
	\includegraphics[width=.9\columnwidth]{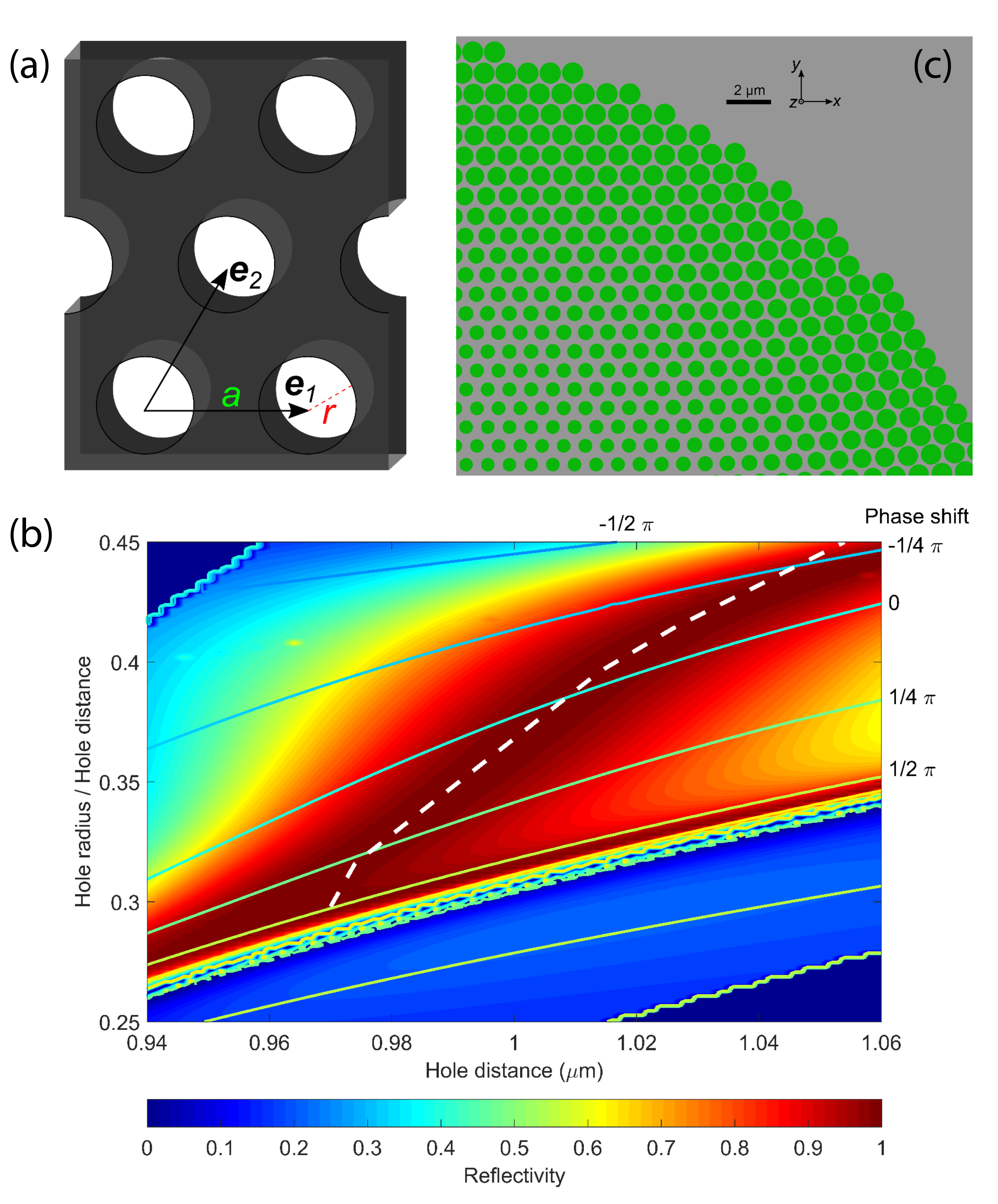}
	\caption{(a) Photonic crystal reflector. Air holes are repeated in the direction of $\vec e_1$ and $\vec e_2$, forming a hexagonal lattice. (b) Reflectivity and phase simulated using RCWA. A phase shift of 0.7$\pi$ is covered by an area with high reflectivity (reflectance $>$98\%). In order to find the best design for our focusing PC, we chose the path highlighted with the white dashed line. The reflectivity in the deep blue area is not calculated, as it is $<$20\%. (c) Final pattern constituting a focusing photonic crystal at 1064~nm.}
	\label{Fig:PCDesign}
\end{figure}

We start with an unmodulated PC, whose thickness and materials are determined by our optomechanical force-sensor design~\cite{Norte2016} and ease of microfabrication. To fully define the basic unit cell we therefore have two parameters left that can be varied to match the desired phase shift and hence form a parameter space from which the whole pattern is generated. In principle, more degrees of freedom can be involved, but only at a significant increase in computational cost. We choose the air holes in our silicon nitride slab to form a hexagonal lattice as the basic shape (see Fig.~\ref{Fig:PCDesign}), where the variables are distance between holes $a$ and filling factor, which is defined as the ratio of air hole radius to hole distance, $\zeta=\frac{r}{a}$. Then, using RCWA, we calculate the reflection amplitude of periodic PC in the parameter space, $\mathcal{R}(a,\zeta)$, giving the reflectivity and phase shift. The corresponding unit cells are building blocks of our focusing PC. We assume that $\mathcal{R}(a,\zeta)$ is equal to the local reflection amplitude on the focusing PC, where a cell with parameters $(a,\zeta)$ is located. Because both the target phase profile and the geometry of the structure vary adiabatically, these parameter values are taken from a smooth and continuous curve. Hence, the two parameters can be described by a single parameter, $(a,\zeta) = (a(p),\zeta(p))$, where $p$ can be viewed as resulting from the parameterization of the curve.

Building a high finesse cavity to maximize the optical read-out sensitivity requires minimizing the optical losses, making a high reflectivity and large numerical aperture of the PC desirable. The latter is a consequence of the diffraction limit and is only important when the full range of a $2\pi$ phase shift is not achievable. Therefore, we should restrict $(a,\zeta)$ to being in an area of high reflectivity, while simultaneously maximizing the achievable phase shift they cover. In addition, considering the achievable accuracy in lithography, small fabrication  errors should have a negligible impact on the phase profile. The procedure outlined above is repeated and optimized, where we vary the thickness and base shape of the design, until the requirements are fulfilled. The main challenges here are the moderate refractive index of SiN ($n = 2$), and the requirement of the membrane to be suspended, in order to achieve high mechanical quality factors. Note that previous devices with multi-layered materials~\cite{Wang2015a,Wang2015b} do not suffer from these limitations.

With a membrane thickness of 325~nm and for a laser wavelength of 1064~nm, we reach a maximum phase change of $\sim0.8\pi$, and a continuous reflectivity of $>$86\% (cf.\ Fig.~\ref{Fig:PCDesign}(b)). This, however, does not give an upperbound for the reflectivity of the photonic crystal as a whole. Considering that the incident beam is a Gaussian wave, there is significantly more power concentrated in the center of the focusing PC, making the reflectance of the cells with smaller phase shift more important. Also, the reflectivity at the edge of the structure, which need large phase shift, suffers more from the edges of the structure, so the reflectance of the corresponding unit cell is less crucial. Hence, with our design, a reflectivity of $>$98\% covers a phase change of $0.7\pi$, which is roughly the maximum achievable overall reflectivity. Also, increasing the size of the PC in general improves the reflectivity~\cite{Chen2017}.

Designing the full device now requires one last step:\ to put the cells into a plane layer. For that, we first specifiy the center cell. If all the chosen parameters are not able to cover a phase shift of full $2\pi$ range, which is the case here, the phase profile of the center cell should be a maximum, which is an ending point of the curve. Then, all other cells are determined. In two dimension, we can label each cell by two numbers, $(n,m)$. By matching the phase
\begin{equation}\label{Eq:PhaseMatch}
\phi_{\mathrm{tag}}(\vec x_{n,m}) = \phi(\vec x_{n,m}) = \mathrm{arg}(\mathcal{R}(a_{n,m},\zeta_{n,m})),
\end{equation}
a focusing PC is obtained. For an unmodulated PC, the location of a cell is $\vec x_{n,m} = n \vec e_1 + m \vec e_2$, where $\vec e_i$ are the primitive lattice vectors ($i=1,2$), and we choose $\vec e_1$ to be parallel to the $x$ axis. In a modulated structure, $\vec e_i$ however is not well defined, since the photonic lattice is distorted. Still, considering that the structure is varied smoothly, $\vec x_{n,m}$ can be approximated as a local vector using neighboring cells. Considering that the direction of $\vec e_i$ should not change much, we use the scheme
\begin{equation}
\begin{gathered}
x_{n,m} = x_{n,m-1} + \frac{a_{n-1,m}+a_{n,m}}{2},\\
y_{n,m} = y_{n-1,m} + \frac{\sqrt 3}{2}\frac{a_{n,m-1}+a_{n,m}}{2},
\end{gathered}
\end{equation}
where the factors $1/2$ and $\sqrt{3}/2$ from the triangular lattice are retained. Since $a_{n,m}=a(p_{n,m})$, combining with equation~(\ref{Eq:PhaseMatch}), $p_{n,m}$ and all undefined parameters for cell $(n,m)$ can be solved, given that cell $(n-1,m)$ and $(n,m-1)$ are already known. Therefore, all the cells can be determined in a systematic iterative process. In general, to determine one cell, two neighboring cells are related. At the beginning, only one cell is predefined. To solve this, the cells on the $x$ axis are found first. This then reduces to a scalar problem, while $y = 0$ is always set. Due to the radial symmetry of the target phase and because the $x$ axis is a lattice axis, fixing these cells on the $x$ axis should not induce any error. We also predefine the cells along another lattice axis. Then, we settle the remaining cells. Part of the generated pattern is shown in Fig.~\ref{Fig:PCDesign}(c), with a total radius of the PC of 30~$\mu$m and a design focal length of 1000~$\mu$m. Far away from the center, both the hole radius and distance are larger. Our final focusing PC design involves a change of $a$ between 0.97 to 1.05~$\mu$m, while we sweep $r/a$ from 0.3 to 0.45. The exact path in this parameter space is shown in Fig.~\ref{Fig:PCDesign}(b).

\begin{figure}[t]
	\centering
	\includegraphics[width=.9\columnwidth]{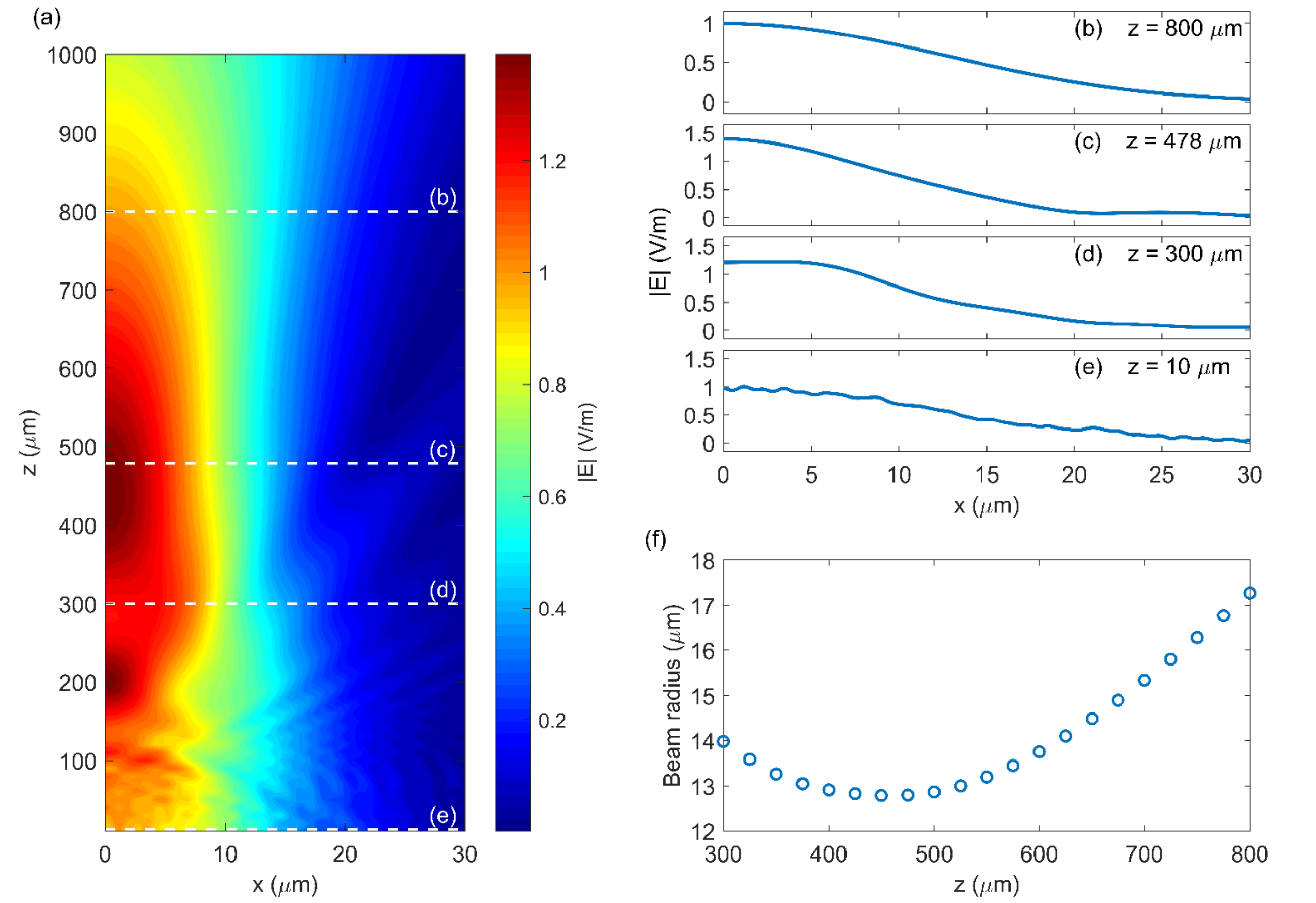}
	\caption{(a) Electric field distribution of the reflected wave assuming an incoming Gaussian beam. The field is maximal between 400--500~$\mu$m distance. Electric field at different distance is shown in (b)-(e). Due to the finite size of the reflector, the reflected beam deviates from an ideal Gaussian distribution for large $x$. (f) Spot radius at different positions, extracted through Gaussian fitting. The minimum of the focused beam can be found around 475~$\mu$m, as expected from theory.}
	\label{Fig:SimResult}
\end{figure}

We finally simulate the full structure with a finite-difference time-domain method solver (FDTD, Lumerical) and the main result is shown in Fig.~\ref{Fig:SimResult}. Note that in general our simulations are done for TE polarized light. We however also verify that an optical TM mode yields the same results due to the symmetry of the PC. With an incoming Gaussian wave with radius of 18~$\mu$m, we get a reflectivity of 96\%, which is close to what we expect from the choice of cells, and a focusing efficiency of 98\%~\cite{Arbabi2015}. The slightly lower reflectivity is primarily due to the power loss at the edge of the reflector. This can be be improved by increasing the size of the total PC -- for example, by increasing the radius of 30~$\mu$m to 45~$\mu$m, the reflectivity increases to 97.5\%. The main limitation here is the large increase in computational costs, as well as the desire to design a PC that fits onto our SiN tethered membrane oscillator, which has a lateral size of about 100~$\mu$m. The particular choice of the beam waist is the optimum for the size of the PC, as a smaller waist would result in an increase in losses as it would sample less of the PC holes. With the beam waist centered on the reflector, the position at which the reflected wave is focused to a minimum is expected to be $z_0 = 478~\mu$m, given by~\cite{Saleh1991}
\begin{equation}
z_0=\frac{f}{1+(f/z_R)^2}.
\end{equation}
Here $z_R = \pi w_0^2 / \lambda$ is the Rayleigh range, and $w_0$ is the beam waist radius of the incoming wave. The deviation of $z_0$ from $f$ is a consequence of the long focal length and small beam waist. We are not able to simulate the electric field distribution as a function of distance directly in the FDTD solver because of its long $f$ and the resulting requirement in memory usage for the computation. Instead, we use Kirchhoff's diffraction formula~\cite{Guenther1990} to generate Fig.~\ref{Fig:SimResult}(a)
\begin{widetext}
\begin{equation}\label{Eq:KDF}
\vec E(\vec x) = - \frac{1}{4\pi} \iint_{\mathrm{reflector}} \mathrm{d}S^\prime
\frac{\exp(i \vec k \cdot (\vec x^\prime - \vec x))}{|\vec x^\prime - \vec x|}	\left[ i k_z \vec E(\vec x^\prime) + \frac{\partial \vec E(\vec x^\prime)}{\partial z^\prime} \right],
\end{equation}
\end{widetext}
where $\vec x^\prime$ is a point on the reflector over which the integral is perform. We approximate the derivative by putting two monitors with a gap of 76~nm, which is the meshing precision in the $z$ direction. The resulting pattern shows a maximum of the electric field between 400 and 500~$\mu$m away from the reflector. Far away from the optical axis, the electric field distribution cannot be approximated by a Gaussian, because of the finite size of the reflector. We fit the electric field for $x \leq 15~\mu$m using a Gaussian function to obtain the spot radius ($1/e^2$ in intensity, cf.\ Fig.~\ref{Fig:SimResult}(b)-(e)) at different distances (Fig.~\ref{Fig:SimResult}(f)). The minimum of the beam spot is found to be around 475~$\mu$m, close to ideal focusing. In our design process, we observe that each cell contributes to the local reflection amplitude, which is a discretization process. Even though the period of each unit cell, given by the hole distance ranging between 0.96~$\mu$m and 1.06~$\mu$m, is close to the optical wavelength, focusing still behaves as expected as small variations at short distance are averaged out far away from the reflector.

\section{Integrated optical sensors}

\begin{figure}[ht!]
	\centering
	\includegraphics[width=.9\columnwidth]{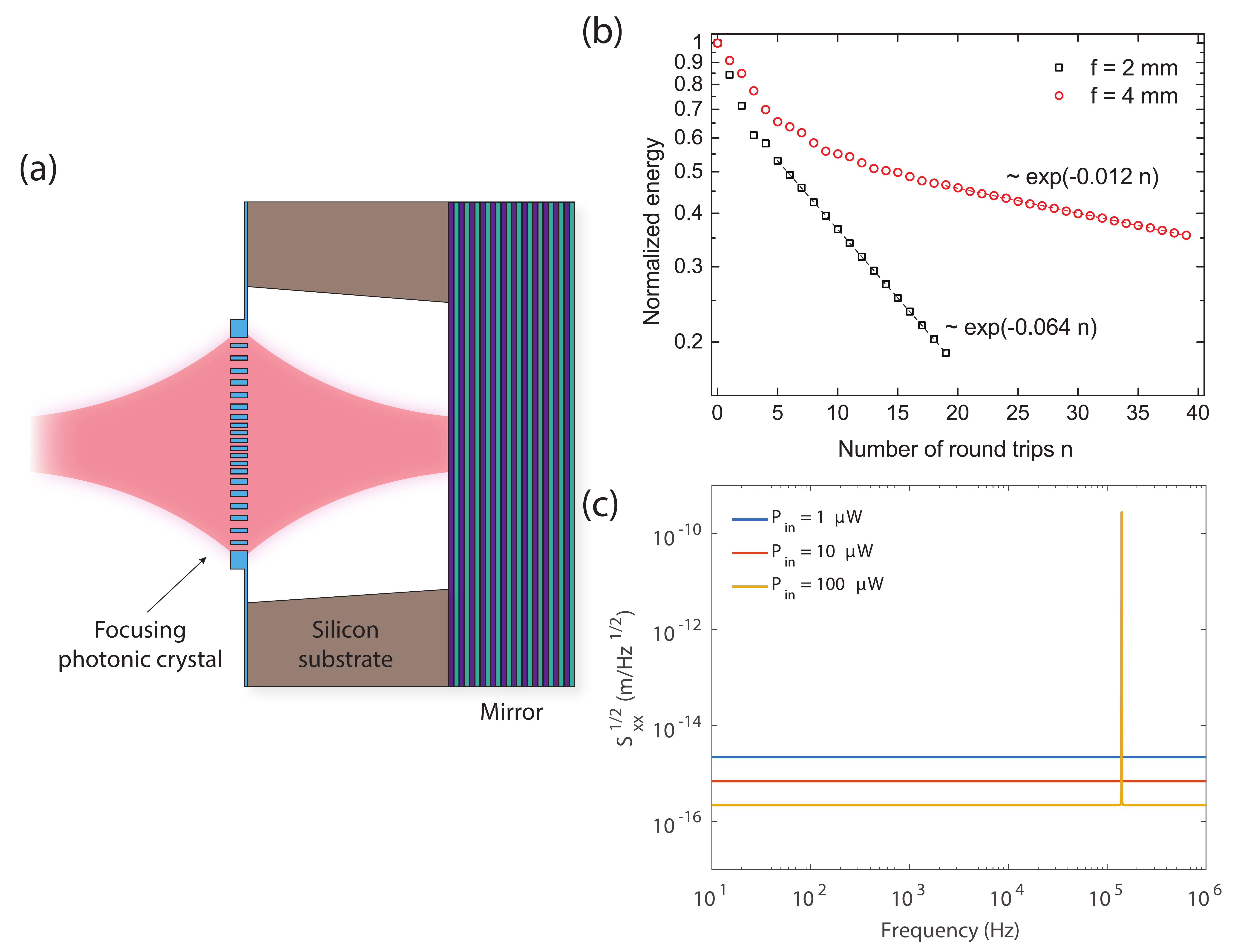}
	\caption{(a) Sketch of a cavity formed by a focusing photonic crystal membrane and a plane mirror. The membrane does not need to be precisely aligned to the center of the mirror, as is usually required for other types of optomechanical cavities. (b) Shown is the field intensity each time the intra-cavity field reaches the PC membrane, normalized to the energy of the initial input plane wave. The initial faster decay is due to higher order modes decaying quickly in the cavity, while the final slope represents the decay rate of the fundamental mode inside the cavity. (c)  Calculated force noise spectrum at low frequency of an optomechanical cavity with a focusing PC membrane, assuming an ideal detector. Already for these very moderate input powers the sensitivity is expected to be well below the standard quantum limit. The bandwidth of such a sensor is limited by the mechanical oscillation frequency, in our case to 140~kHz.}
	\label{Fig:Cavity}
\end{figure}

One potential application for the mirror we designed could be in an optomechanical cavity, which is conventionally made of a highly-reflective membrane and a separate focusing mirror~\cite{Gigan2006,Arcizet2006b,Aspelmeyer2014,Reinhardt2016}. While the mirror is fixed, the mechanical oscillation of the membrane is of great interest. Such a concept has been used to sense acceleration and other forces~\cite{Krause2012,Metcalfe2014,deLepinay2016}. To achieve high sensitivity, the optomechanical cavity should have a high optical quality factor. However, as the reflective membrane is usually small, any slight displacement of the membrane from the optical axis of the mirror greatly increases the loss, leading to a tedious alignment process usually involving sophisticated motorized stages. Such setups are bulky, complicated, expensive and inherently susceptible to vibrations. With our design method, we could use the focusing PC as an optomechanical cavity membrane, while at the same time focusing the optical beam onto a fixed, highly reflective mirror, that now can be part of a monolithic cavity (see Fig.~\ref{Fig:Cavity}(a)). The fixed mirror could for example be directly integrated as a distributed Bragg reflector, as already commonly used on chips in optomechanics experiments~\cite{Groeblacher2009a}.

\begin{table}[]
	\centering
	\begin{tabular}{|c|c|c|c|}
		\hline
		Focal length & PC radius & Cavity length & Finesse \\\hline
		1~mm & 29~$\mu$m & 475~$\mu$m & 56 \\\hline
		2~mm & 41~$\mu$m & 1~mm & 57 \\\hline
		4~mm & 58~$\mu$m & 1~mm & 103 \\\hline
	\end{tabular}
	\caption{Performance of various focusing photonic crystal cavities assuming a fixed mirror with 99\% reflectivity. All photonic crystal designs exhibit an average reflectivity of 96\%, while the optical phase shift is around 0.8$\pi$.}
	\label{Tab:RefPerf}
\end{table}

One of the main limitation in our design are the cavity losses that result from the finite overlap of the optical mode and the membrane. Using equation~(\ref{Eq:KDF}) we can trace the intracavity field and, assuming a perfectly reflective membrane, calculate the resulting intensity decay rate. With a cavity length of 1~mm and focal length 2~mm, the loss per round trip is 6.4\%, while for a focal length of 4~mm the losses can be reduced to 1.2\% (Fig.~\ref{Fig:Cavity}(b)). Further increase of the focal length can result in even less losses and at the same time raises the membrane mass, while it adds difficulties in fabrication beyond practical limits due to the size of the membrane. With a phase shift of $0.8\pi$, the radius of the reflector is 58~$\mu$m for a 4~mm focal length (see table~\ref{Tab:RefPerf}). We will concentrate on this design, as the losses are already smaller than the loss on the membrane. Combining the two losses, the lower bound for the cavity finesse is estimated to be 103 (corresponding to a quality factor of $3.6\times 10^4$). We can now combine this design with our previously fabricated ultra-low mechanical loss membranes~\cite{Norte2016}. By modifying the pattern of the original reflector, a focusing PC can be realized in an otherwise almost unchanged silicon nitride membrane, resulting in a similar mechanical quality factor. We now use such a device as a model to analyze the performance of such a cavity sensor. Considering that the thickness is increased from 20~nm to 325~nm, the membrane is heavier, slightly reducing the sensitivity. Therefore, the intrinsic thermal noise of the mechanical oscillator is approx.\ $4.4 \times 10^{-17}$~m/Hz$^{-1/2}$ at room temperature. In this cavity, the second, fixed mirror is assumed to be flat with a reflectivity of 99\%. We calculate an achievable displacement sensitivity of $7 \times 10^{-16}$~m/Hz$^{1/2}$ at 10~$\mu$W, around one order of magnitude smaller than the standard quantum limit, while the force noise of the measurement can be expected to be below $8.7 \times 10^{-15}$~N/Hz$^{1/2}$, if the transmitted power is measured by an ideal detector. With the modified mass being 16~ng, such a sensor would correspond to an accelerometer with a sensitivity of 54.6~$\mu$g/Hz$^{1/2}$, comparable to other recent optomechanics sensors~\cite{Krause2012}.

\section{Conclusion}
In summary, we describe a simple way of designing a stable, self-aligning optomechanical cavity, that should be able to have state-of-the-art force sensitivity. This is thanks to a novel single dielectric design using highly-stressed silicon nitride membranes, which are known to exhibit exceptional mechanical properties. We believe that such an integrated cavity could find applications both in precision measurements as well as in the field of optomechanics.

\section{Acknowledgements}
We would like to thank Jo\~{a}o Moura for helpful discussions. This project was supported by the European Research Council (ERC StG Strong-Q, grant agreement 676842) and by the Netherlands Organisation for Scientific Research (NWO/OCW), as part of the Frontiers of Nanoscience program.


\begin{thebibliography}{33}
	\expandafter\ifx\csname natexlab\endcsname\relax\def\natexlab#1{#1}\fi
	\expandafter\ifx\csname bibnamefont\endcsname\relax
	\def\bibnamefont#1{#1}\fi
	\expandafter\ifx\csname bibfnamefont\endcsname\relax
	\def\bibfnamefont#1{#1}\fi
	\expandafter\ifx\csname citenamefont\endcsname\relax
	\def\citenamefont#1{#1}\fi
	\expandafter\ifx\csname url\endcsname\relax
	\def\url#1{\texttt{#1}}\fi
	\expandafter\ifx\csname urlprefix\endcsname\relax\def\urlprefix{URL }\fi
	\providecommand{\bibinfo}[2]{#2}
	\providecommand{\eprint}[2][]{\url{#2}}

	\bibitem[{\citenamefont{Boutami et~al.}(2006)\citenamefont{Boutami, Bakir,
			Leclercq, Letartre, Rojo-Romeo, Garrigues, Viktorovitch, Sagnes, Legratiet,
			and Strassner}}]{Boutami2006}
	\bibinfo{author}{\bibfnamefont{S.}~\bibnamefont{Boutami}},
	\bibinfo{author}{\bibfnamefont{B.~B.} \bibnamefont{Bakir}},
	\bibinfo{author}{\bibfnamefont{J.-L.} \bibnamefont{Leclercq}},
	\bibinfo{author}{\bibfnamefont{X.}~\bibnamefont{Letartre}},
	\bibinfo{author}{\bibfnamefont{P.}~\bibnamefont{Rojo-Romeo}},
	\bibinfo{author}{\bibfnamefont{M.}~\bibnamefont{Garrigues}},
	\bibinfo{author}{\bibfnamefont{P.}~\bibnamefont{Viktorovitch}},
	\bibinfo{author}{\bibfnamefont{I.}~\bibnamefont{Sagnes}},
	\bibinfo{author}{\bibfnamefont{L.}~\bibnamefont{Legratiet}},
	\bibnamefont{and}
	\bibinfo{author}{\bibfnamefont{M.}~\bibnamefont{Strassner}},
	\bibinfo{journal}{Opt.\ Express} \textbf{\bibinfo{volume}{14}},
	\bibinfo{pages}{3129} (\bibinfo{year}{2006}).

	\bibitem[{\citenamefont{Wang et~al.}(2015{\natexlab{a}})\citenamefont{Wang,
			Stellinga, Klemm, Reardon, and Krauss}}]{Wang2015a}
	\bibinfo{author}{\bibfnamefont{Y.}~\bibnamefont{Wang}},
	\bibinfo{author}{\bibfnamefont{D.}~\bibnamefont{Stellinga}},
	\bibinfo{author}{\bibfnamefont{A.~B.} \bibnamefont{Klemm}},
	\bibinfo{author}{\bibfnamefont{C.~P.} \bibnamefont{Reardon}},
	\bibnamefont{and} \bibinfo{author}{\bibfnamefont{T.~F.}
		\bibnamefont{Krauss}}, \bibinfo{journal}{IEEE J.\ Sel.\ Top.\ Quantum
		Electron.} \textbf{\bibinfo{volume}{21}}, \bibinfo{pages}{2700706}
	(\bibinfo{year}{2015}{\natexlab{a}}).

	\bibitem[{\citenamefont{Wang et~al.}(2015{\natexlab{b}})\citenamefont{Wang,
			Gao, Fang, Li, Zhu, and Wang}}]{Wang2015b}
	\bibinfo{author}{\bibfnamefont{W.}~\bibnamefont{Wang}},
	\bibinfo{author}{\bibfnamefont{X.}~\bibnamefont{Gao}},
	\bibinfo{author}{\bibfnamefont{X.}~\bibnamefont{Fang}},
	\bibinfo{author}{\bibfnamefont{X.}~\bibnamefont{Li}},
	\bibinfo{author}{\bibfnamefont{H.}~\bibnamefont{Zhu}}, \bibnamefont{and}
	\bibinfo{author}{\bibfnamefont{Y.}~\bibnamefont{Wang}},
	\bibinfo{journal}{IEEE Photon.\ J.} \textbf{\bibinfo{volume}{8}},
	\bibinfo{pages}{6800614} (\bibinfo{year}{2015}{\natexlab{b}}).

	\bibitem[{\citenamefont{Li et~al.}(2011)\citenamefont{Li, Fattal, Fiorentino,
			and Beausoleil}}]{Li2011}
	\bibinfo{author}{\bibfnamefont{J.}~\bibnamefont{Li}},
	\bibinfo{author}{\bibfnamefont{D.}~\bibnamefont{Fattal}},
	\bibinfo{author}{\bibfnamefont{M.}~\bibnamefont{Fiorentino}},
	\bibnamefont{and} \bibinfo{author}{\bibfnamefont{R.~G.}
		\bibnamefont{Beausoleil}}, \bibinfo{journal}{Phys.\ Rev.\ Lett.}
	\textbf{\bibinfo{volume}{106}}, \bibinfo{pages}{193901}
	(\bibinfo{year}{2011}).

	\bibitem[{\citenamefont{Feng et~al.}(2005)\citenamefont{Feng, Li, Feng, Ren,
			Cheng, and Zhang}}]{Feng2005}
	\bibinfo{author}{\bibfnamefont{S.}~\bibnamefont{Feng}},
	\bibinfo{author}{\bibfnamefont{Z.-Y.} \bibnamefont{Li}},
	\bibinfo{author}{\bibfnamefont{Z.-F.} \bibnamefont{Feng}},
	\bibinfo{author}{\bibfnamefont{K.}~\bibnamefont{Ren}},
	\bibinfo{author}{\bibfnamefont{B.-Y.} \bibnamefont{Cheng}}, \bibnamefont{and}
	\bibinfo{author}{\bibfnamefont{D.-Z.} \bibnamefont{Zhang}},
	\bibinfo{journal}{J.\ Appl.\ Phys.} \textbf{\bibinfo{volume}{98}},
	\bibinfo{pages}{063102} (\bibinfo{year}{2005}).

	\bibitem[{\citenamefont{Fattal et~al.}(2010)\citenamefont{Fattal, Li, Peng,
			Fiorentino, and Beausoleil}}]{Fattal2010}
	\bibinfo{author}{\bibfnamefont{D.}~\bibnamefont{Fattal}},
	\bibinfo{author}{\bibfnamefont{J.}~\bibnamefont{Li}},
	\bibinfo{author}{\bibfnamefont{Z.}~\bibnamefont{Peng}},
	\bibinfo{author}{\bibfnamefont{M.}~\bibnamefont{Fiorentino}},
	\bibnamefont{and} \bibinfo{author}{\bibfnamefont{R.~G.}
		\bibnamefont{Beausoleil}}, \bibinfo{journal}{Nature Photon.}
	\textbf{\bibinfo{volume}{4}}, \bibinfo{pages}{466} (\bibinfo{year}{2010}).

	\bibitem[{\citenamefont{Lu et~al.}(2010)\citenamefont{Lu, Sedgwick, Karagodsky,
			Chase, and Chang-Hasnain}}]{Lu2010}
	\bibinfo{author}{\bibfnamefont{F.}~\bibnamefont{Lu}},
	\bibinfo{author}{\bibfnamefont{F.~G.} \bibnamefont{Sedgwick}},
	\bibinfo{author}{\bibfnamefont{V.}~\bibnamefont{Karagodsky}},
	\bibinfo{author}{\bibfnamefont{C.}~\bibnamefont{Chase}}, \bibnamefont{and}
	\bibinfo{author}{\bibfnamefont{C.~J.} \bibnamefont{Chang-Hasnain}},
	\bibinfo{journal}{Opt.\ Express} \textbf{\bibinfo{volume}{18}},
	\bibinfo{pages}{12606} (\bibinfo{year}{2010}).

	\bibitem[{\citenamefont{Zhan et~al.}(2016)\citenamefont{Zhan, Colburn, Trivedi,
			Fryett, Dodson, and Majumdar}}]{Zhan2016}
	\bibinfo{author}{\bibfnamefont{A.}~\bibnamefont{Zhan}},
	\bibinfo{author}{\bibfnamefont{S.}~\bibnamefont{Colburn}},
	\bibinfo{author}{\bibfnamefont{R.}~\bibnamefont{Trivedi}},
	\bibinfo{author}{\bibfnamefont{T.~K.} \bibnamefont{Fryett}},
	\bibinfo{author}{\bibfnamefont{C.~M.} \bibnamefont{Dodson}},
	\bibnamefont{and} \bibinfo{author}{\bibfnamefont{A.}~\bibnamefont{Majumdar}},
	\bibinfo{journal}{ACS Photonics} \textbf{\bibinfo{volume}{3}},
	\bibinfo{pages}{209} (\bibinfo{year}{2016}).

	\bibitem[{\citenamefont{Letartre et~al.}(2003)\citenamefont{Letartre, Mouette,
			Leclercq, Romeo, Seassal, and Viktorovitch}}]{Letartre2003}
	\bibinfo{author}{\bibfnamefont{X.}~\bibnamefont{Letartre}},
	\bibinfo{author}{\bibfnamefont{J.}~\bibnamefont{Mouette}},
	\bibinfo{author}{\bibfnamefont{J.~L.} \bibnamefont{Leclercq}},
	\bibinfo{author}{\bibfnamefont{P.~R.} \bibnamefont{Romeo}},
	\bibinfo{author}{\bibfnamefont{C.}~\bibnamefont{Seassal}}, \bibnamefont{and}
	\bibinfo{author}{\bibfnamefont{P.}~\bibnamefont{Viktorovitch}},
	\bibinfo{journal}{J.\ Lightwave Technol.} \textbf{\bibinfo{volume}{21}}
	(\bibinfo{year}{2003}).

	\bibitem[{\citenamefont{Mateus et~al.}(2004)\citenamefont{Mateus, Huang, Chen,
			Chang-Hasnain, and Suzuki}}]{Mateus2004}
	\bibinfo{author}{\bibfnamefont{C.~F.~R.} \bibnamefont{Mateus}},
	\bibinfo{author}{\bibfnamefont{M.~C.~Y.} \bibnamefont{Huang}},
	\bibinfo{author}{\bibfnamefont{L.}~\bibnamefont{Chen}},
	\bibinfo{author}{\bibfnamefont{C.~J.} \bibnamefont{Chang-Hasnain}},
	\bibnamefont{and} \bibinfo{author}{\bibfnamefont{Y.}~\bibnamefont{Suzuki}},
	\bibinfo{journal}{IEEE Photon.\ Technol.\ Lett.}
	\textbf{\bibinfo{volume}{16}}, \bibinfo{pages}{1676} (\bibinfo{year}{2004}).

	\bibitem[{\citenamefont{Kamali et~al.}(2016)\citenamefont{Kamali, Arbabi,
			Arbabi, Horie, and Faraon}}]{Kamali2016}
	\bibinfo{author}{\bibfnamefont{S.~M.} \bibnamefont{Kamali}},
	\bibinfo{author}{\bibfnamefont{E.}~\bibnamefont{Arbabi}},
	\bibinfo{author}{\bibfnamefont{A.}~\bibnamefont{Arbabi}},
	\bibinfo{author}{\bibfnamefont{Y.}~\bibnamefont{Horie}}, \bibnamefont{and}
	\bibinfo{author}{\bibfnamefont{A.}~\bibnamefont{Faraon}},
	\bibinfo{journal}{Laser Photonics Rev.} \textbf{\bibinfo{volume}{10}},
	\bibinfo{pages}{1002} (\bibinfo{year}{2016}).

	\bibitem[{\citenamefont{Stambaugh et~al.}(2015)\citenamefont{Stambaugh, Xu,
			Kemiktarak, Taylor, and Lawall}}]{Stambaugh2015}
	\bibinfo{author}{\bibfnamefont{C.}~\bibnamefont{Stambaugh}},
	\bibinfo{author}{\bibfnamefont{H.}~\bibnamefont{Xu}},
	\bibinfo{author}{\bibfnamefont{U.}~\bibnamefont{Kemiktarak}},
	\bibinfo{author}{\bibfnamefont{J.}~\bibnamefont{Taylor}}, \bibnamefont{and}
	\bibinfo{author}{\bibfnamefont{J.}~\bibnamefont{Lawall}},
	\bibinfo{journal}{Ann.\ Phys.} \textbf{\bibinfo{volume}{527}},
	\bibinfo{pages}{81} (\bibinfo{year}{2015}).

	\bibitem[{\citenamefont{Crozier et~al.}(2006)\citenamefont{Crozier, Lousse,
			Kilic, Kim, Fan, and Solgaard}}]{Crozier2006}
	\bibinfo{author}{\bibfnamefont{K.~B.} \bibnamefont{Crozier}},
	\bibinfo{author}{\bibfnamefont{V.}~\bibnamefont{Lousse}},
	\bibinfo{author}{\bibfnamefont{O.}~\bibnamefont{Kilic}},
	\bibinfo{author}{\bibfnamefont{S.}~\bibnamefont{Kim}},
	\bibinfo{author}{\bibfnamefont{S.}~\bibnamefont{Fan}}, \bibnamefont{and}
	\bibinfo{author}{\bibfnamefont{O.}~\bibnamefont{Solgaard}},
	\bibinfo{journal}{Phys.\ Rev.\ B} \textbf{\bibinfo{volume}{73}},
	\bibinfo{pages}{115126} (\bibinfo{year}{2006}).

	\bibitem[{\citenamefont{Bui et~al.}(2012)\citenamefont{Bui, Zheng, Hoch, Lee,
			Harris, and Wong}}]{Bui2012}
	\bibinfo{author}{\bibfnamefont{C.~H.} \bibnamefont{Bui}},
	\bibinfo{author}{\bibfnamefont{J.}~\bibnamefont{Zheng}},
	\bibinfo{author}{\bibfnamefont{S.~W.} \bibnamefont{Hoch}},
	\bibinfo{author}{\bibfnamefont{L.~Y.~T.} \bibnamefont{Lee}},
	\bibinfo{author}{\bibfnamefont{J.~G.~E.} \bibnamefont{Harris}},
	\bibnamefont{and} \bibinfo{author}{\bibfnamefont{C.~W.} \bibnamefont{Wong}},
	\bibinfo{journal}{Appl.\ Phys.\ Lett.} \textbf{\bibinfo{volume}{100}},
	\bibinfo{pages}{021110} (\bibinfo{year}{2012}).

	\bibitem[{\citenamefont{Norte et~al.}(2016)\citenamefont{Norte, Moura, and
			Gr\"oblacher}}]{Norte2016}
	\bibinfo{author}{\bibfnamefont{R.~A.} \bibnamefont{Norte}},
	\bibinfo{author}{\bibfnamefont{J.~P.} \bibnamefont{Moura}}, \bibnamefont{and}
	\bibinfo{author}{\bibfnamefont{S.}~\bibnamefont{Gr\"oblacher}},
	\bibinfo{journal}{Phys.\ Rev.\ Lett.} \textbf{\bibinfo{volume}{116}},
	\bibinfo{pages}{147202} (\bibinfo{year}{2016}).

	\bibitem[{\citenamefont{Bernard et~al.}(2016)\citenamefont{Bernard, Reinhardt,
			Dumont, Peter, and Sankey}}]{Bernard2016}
	\bibinfo{author}{\bibfnamefont{S.}~\bibnamefont{Bernard}},
	\bibinfo{author}{\bibfnamefont{C.}~\bibnamefont{Reinhardt}},
	\bibinfo{author}{\bibfnamefont{V.}~\bibnamefont{Dumont}},
	\bibinfo{author}{\bibfnamefont{Y.-A.} \bibnamefont{Peter}}, \bibnamefont{and}
	\bibinfo{author}{\bibfnamefont{J.~C.} \bibnamefont{Sankey}},
	\bibinfo{journal}{Opt.\ Lett.} \textbf{\bibinfo{volume}{41}},
	\bibinfo{pages}{5624} (\bibinfo{year}{2016}).

	\bibitem[{\citenamefont{Chen et~al.}(2017)\citenamefont{Chen, Chardin, Makles,
			Ca\"{e}r, Chua, Braive, Robert-Philip, Briant, Cohadon, Heidmann
			et~al.}}]{Chen2017}
	\bibinfo{author}{\bibfnamefont{X.}~\bibnamefont{Chen}},
	\bibinfo{author}{\bibfnamefont{C.}~\bibnamefont{Chardin}},
	\bibinfo{author}{\bibfnamefont{K.}~\bibnamefont{Makles}},
	\bibinfo{author}{\bibfnamefont{C.}~\bibnamefont{Ca\"{e}r}},
	\bibinfo{author}{\bibfnamefont{S.}~\bibnamefont{Chua}},
	\bibinfo{author}{\bibfnamefont{R.}~\bibnamefont{Braive}},
	\bibinfo{author}{\bibfnamefont{I.}~\bibnamefont{Robert-Philip}},
	\bibinfo{author}{\bibfnamefont{T.}~\bibnamefont{Briant}},
	\bibinfo{author}{\bibfnamefont{P.-F.} \bibnamefont{Cohadon}},
	\bibinfo{author}{\bibfnamefont{A.}~\bibnamefont{Heidmann}},
	\bibnamefont{et~al.}, \bibinfo{journal}{Light Sci.\ Appl.}
	\textbf{\bibinfo{volume}{6}}, \bibinfo{pages}{e16190} (\bibinfo{year}{2017}).

	\bibitem[{\citenamefont{Xuereb et~al.}(2012)\citenamefont{Xuereb, Genes, and
			Dantan}}]{Xuereb2012}
	\bibinfo{author}{\bibfnamefont{A.}~\bibnamefont{Xuereb}},
	\bibinfo{author}{\bibfnamefont{C.}~\bibnamefont{Genes}}, \bibnamefont{and}
	\bibinfo{author}{\bibfnamefont{A.}~\bibnamefont{Dantan}},
	\bibinfo{journal}{Phys.\ Rev.\ Lett.} \textbf{\bibinfo{volume}{109}},
	\bibinfo{pages}{223601} (\bibinfo{year}{2012}).

	\bibitem[{\citenamefont{Joannopoulos et~al.}(2008)\citenamefont{Joannopoulos,
			Johnson, Winn, and Meade}}]{Joannopoulos2008}
	\bibinfo{author}{\bibfnamefont{J.~D.} \bibnamefont{Joannopoulos}},
	\bibinfo{author}{\bibfnamefont{S.~G.} \bibnamefont{Johnson}},
	\bibinfo{author}{\bibfnamefont{J.~N.} \bibnamefont{Winn}}, \bibnamefont{and}
	\bibinfo{author}{\bibfnamefont{R.~D.} \bibnamefont{Meade}},
	\emph{\bibinfo{title}{Photonic Crystals: Molding the Flow of Light (Second
			Edition)}} (\bibinfo{publisher}{Princeton University}, \bibinfo{year}{2008}),
	\bibinfo{edition}{2nd} ed.

	\bibitem[{\citenamefont{Moharam and Gaylord}(1981)}]{Moharam1981}
	\bibinfo{author}{\bibfnamefont{M.~G.} \bibnamefont{Moharam}} \bibnamefont{and}
	\bibinfo{author}{\bibfnamefont{T.~K.} \bibnamefont{Gaylord}},
	\bibinfo{journal}{J.\ Opt.\ Soc.\ Am.} \textbf{\bibinfo{volume}{71}},
	\bibinfo{pages}{811} (\bibinfo{year}{1981}).

	\bibitem[{\citenamefont{Rumpf}(2011)}]{Rumpf2011}
	\bibinfo{author}{\bibfnamefont{R.~C.} \bibnamefont{Rumpf}},
	\bibinfo{journal}{Prog.\ Electromagn.\ Res.\ B}
	\textbf{\bibinfo{volume}{35}}, \bibinfo{pages}{241} (\bibinfo{year}{2011}).

	\bibitem[{\citenamefont{Klemm et~al.}(2013)\citenamefont{Klemm, Stellinga,
			Martins, Lewis, Huyet, O'Faolain, and Krauss}}]{Klemm2013}
	\bibinfo{author}{\bibfnamefont{A.~B.} \bibnamefont{Klemm}},
	\bibinfo{author}{\bibfnamefont{D.}~\bibnamefont{Stellinga}},
	\bibinfo{author}{\bibfnamefont{E.~R.} \bibnamefont{Martins}},
	\bibinfo{author}{\bibfnamefont{L.}~\bibnamefont{Lewis}},
	\bibinfo{author}{\bibfnamefont{G.}~\bibnamefont{Huyet}},
	\bibinfo{author}{\bibfnamefont{L.}~\bibnamefont{O'Faolain}},
	\bibnamefont{and} \bibinfo{author}{\bibfnamefont{T.~F.}
		\bibnamefont{Krauss}}, \bibinfo{journal}{Opt.\ Lett.}
	\textbf{\bibinfo{volume}{38}}, \bibinfo{pages}{3410} (\bibinfo{year}{2013}).

	\bibitem[{\citenamefont{Arbabi et~al.}(2015)\citenamefont{Arbabi, Horie, Ball,
			Bagheri, and Faraon}}]{Arbabi2015}
	\bibinfo{author}{\bibfnamefont{A.}~\bibnamefont{Arbabi}},
	\bibinfo{author}{\bibfnamefont{Y.}~\bibnamefont{Horie}},
	\bibinfo{author}{\bibfnamefont{A.~J.} \bibnamefont{Ball}},
	\bibinfo{author}{\bibfnamefont{M.}~\bibnamefont{Bagheri}}, \bibnamefont{and}
	\bibinfo{author}{\bibfnamefont{A.}~\bibnamefont{Faraon}},
	\bibinfo{journal}{Nature Commun.} \textbf{\bibinfo{volume}{6}},
	\bibinfo{pages}{7069} (\bibinfo{year}{2015}).

	\bibitem[{\citenamefont{Saleh and Teich}(1991)}]{Saleh1991}
	\bibinfo{author}{\bibfnamefont{B.~E.~A.} \bibnamefont{Saleh}} \bibnamefont{and}
	\bibinfo{author}{\bibfnamefont{M.~C.} \bibnamefont{Teich}},
	\emph{\bibinfo{title}{Fundamentals of Photonics}} (\bibinfo{publisher}{John
		Wiley \& Sons, Inc.}, \bibinfo{year}{1991}).

	\bibitem[{\citenamefont{Guenther}(1990)}]{Guenther1990}
	\bibinfo{author}{\bibfnamefont{B.~D.} \bibnamefont{Guenther}},
	\emph{\bibinfo{title}{Modern Optics}} (\bibinfo{publisher}{Wiley},
	\bibinfo{address}{New York}, \bibinfo{year}{1990}).

	\bibitem[{\citenamefont{Gigan et~al.}(2006)\citenamefont{Gigan, B\"ohm,
			Paternostro, Blaser, Langer, Hertzberg, Schwab, B\"auerle, Aspelmeyer, and
			Zeilinger}}]{Gigan2006}
	\bibinfo{author}{\bibfnamefont{S.}~\bibnamefont{Gigan}},
	\bibinfo{author}{\bibfnamefont{H.~R.} \bibnamefont{B\"ohm}},
	\bibinfo{author}{\bibfnamefont{M.}~\bibnamefont{Paternostro}},
	\bibinfo{author}{\bibfnamefont{F.}~\bibnamefont{Blaser}},
	\bibinfo{author}{\bibfnamefont{G.}~\bibnamefont{Langer}},
	\bibinfo{author}{\bibfnamefont{J.~B.} \bibnamefont{Hertzberg}},
	\bibinfo{author}{\bibfnamefont{K.~C.} \bibnamefont{Schwab}},
	\bibinfo{author}{\bibfnamefont{D.}~\bibnamefont{B\"auerle}},
	\bibinfo{author}{\bibfnamefont{M.}~\bibnamefont{Aspelmeyer}},
	\bibnamefont{and}
	\bibinfo{author}{\bibfnamefont{A.}~\bibnamefont{Zeilinger}},
	\bibinfo{journal}{Nature} \textbf{\bibinfo{volume}{444}}, \bibinfo{pages}{67}
	(\bibinfo{year}{2006}).

	\bibitem[{\citenamefont{Arcizet et~al.}(2006)\citenamefont{Arcizet, Cohadon,
			Briant, Pinard, and Heidmann}}]{Arcizet2006b}
	\bibinfo{author}{\bibfnamefont{O.}~\bibnamefont{Arcizet}},
	\bibinfo{author}{\bibfnamefont{P.-F.} \bibnamefont{Cohadon}},
	\bibinfo{author}{\bibfnamefont{T.}~\bibnamefont{Briant}},
	\bibinfo{author}{\bibfnamefont{M.}~\bibnamefont{Pinard}}, \bibnamefont{and}
	\bibinfo{author}{\bibfnamefont{A.}~\bibnamefont{Heidmann}},
	\bibinfo{journal}{Nature} \textbf{\bibinfo{volume}{444}}, \bibinfo{pages}{71}
	(\bibinfo{year}{2006}).

	\bibitem[{\citenamefont{Aspelmeyer et~al.}(2014)\citenamefont{Aspelmeyer,
			Kippenberg, and Marquardt}}]{Aspelmeyer2014}
	\bibinfo{author}{\bibfnamefont{M.}~\bibnamefont{Aspelmeyer}},
	\bibinfo{author}{\bibfnamefont{T.~J.} \bibnamefont{Kippenberg}},
	\bibnamefont{and}
	\bibinfo{author}{\bibfnamefont{F.}~\bibnamefont{Marquardt}},
	\bibinfo{journal}{Rev.\ Mod.\ Phys.} \textbf{\bibinfo{volume}{86}},
	\bibinfo{pages}{1391} (\bibinfo{year}{2014}).

	\bibitem[{\citenamefont{Reinhardt et~al.}(2016)\citenamefont{Reinhardt,
			M\"uller, Bourassa, and Sankey}}]{Reinhardt2016}
	\bibinfo{author}{\bibfnamefont{C.}~\bibnamefont{Reinhardt}},
	\bibinfo{author}{\bibfnamefont{T.}~\bibnamefont{M\"uller}},
	\bibinfo{author}{\bibfnamefont{A.}~\bibnamefont{Bourassa}}, \bibnamefont{and}
	\bibinfo{author}{\bibfnamefont{J.~C.} \bibnamefont{Sankey}},
	\bibinfo{journal}{Phys.\ Rev.\ X} \textbf{\bibinfo{volume}{6}},
	\bibinfo{pages}{021001} (\bibinfo{year}{2016}).

	\bibitem[{\citenamefont{Krause et~al.}(2012)\citenamefont{Krause, Winger,
			Blasius, Lin, and Painter}}]{Krause2012}
	\bibinfo{author}{\bibfnamefont{A.~G.} \bibnamefont{Krause}},
	\bibinfo{author}{\bibfnamefont{M.}~\bibnamefont{Winger}},
	\bibinfo{author}{\bibfnamefont{T.~D.} \bibnamefont{Blasius}},
	\bibinfo{author}{\bibfnamefont{Q.}~\bibnamefont{Lin}}, \bibnamefont{and}
	\bibinfo{author}{\bibfnamefont{O.}~\bibnamefont{Painter}},
	\bibinfo{journal}{Nature Photon.} \textbf{\bibinfo{volume}{6}},
	\bibinfo{pages}{768} (\bibinfo{year}{2012}).

	\bibitem[{\citenamefont{Metcalfe}(2014)}]{Metcalfe2014}
	\bibinfo{author}{\bibfnamefont{M.}~\bibnamefont{Metcalfe}},
	\bibinfo{journal}{Appl.\ Phys.\ Rev.} \textbf{\bibinfo{volume}{1}},
	\bibinfo{pages}{031105} (\bibinfo{year}{2014}).

	\bibitem[{\citenamefont{de~L\'{e}pinay
			et~al.}(2016)\citenamefont{de~L\'{e}pinay, Pigeau, Besga, Vincent, Poncharal,
			and Arcizet}}]{deLepinay2016}
	\bibinfo{author}{\bibfnamefont{L.~M.} \bibnamefont{de~L\'{e}pinay}},
	\bibinfo{author}{\bibfnamefont{B.}~\bibnamefont{Pigeau}},
	\bibinfo{author}{\bibfnamefont{B.}~\bibnamefont{Besga}},
	\bibinfo{author}{\bibfnamefont{P.}~\bibnamefont{Vincent}},
	\bibinfo{author}{\bibfnamefont{P.}~\bibnamefont{Poncharal}},
	\bibnamefont{and} \bibinfo{author}{\bibfnamefont{O.}~\bibnamefont{Arcizet}},
	\bibinfo{journal}{Nature Nanotechnol.} p. \bibinfo{pages}{AOP}
	(\bibinfo{year}{2016}).

	\bibitem[{\citenamefont{Gr\"{o}blacher
			et~al.}(2009)\citenamefont{Gr\"{o}blacher, Hertzberg, Vanner, Gigan, Schwab,
			and Aspelmeyer}}]{Groeblacher2009a}
	\bibinfo{author}{\bibfnamefont{S.}~\bibnamefont{Gr\"{o}blacher}},
	\bibinfo{author}{\bibfnamefont{J.~B.} \bibnamefont{Hertzberg}},
	\bibinfo{author}{\bibfnamefont{M.~R.} \bibnamefont{Vanner}},
	\bibinfo{author}{\bibfnamefont{S.}~\bibnamefont{Gigan}},
	\bibinfo{author}{\bibfnamefont{K.~C.} \bibnamefont{Schwab}},
	\bibnamefont{and}
	\bibinfo{author}{\bibfnamefont{M.}~\bibnamefont{Aspelmeyer}},
	\bibinfo{journal}{Nature Phys.} \textbf{\bibinfo{volume}{5}},
	\bibinfo{pages}{485} (\bibinfo{year}{2009}).

\end{thebibliography}

\end{document}